\documentclass[twocolumn,showpacs,preprintnumbers,amsmath,amssymb]{revtex4}

\usepackage{graphicx}
\usepackage{dcolumn}
\usepackage{bm}

\begin{document}

\preprint{APS/123-QED}

\title{Effect of transient pinning on stability of drops sitting on an inclined plane}

\author{Viatcheslav Berejnov}
\altaffiliation[Also at ]{Institute for Integrated Energy Systems,
University of Victoria, P.O. Box 3055, Victoria, BC V8W 3P6,
Canada} \email{berejnov@uvic.ca}

\author{Robert E. Thorne}

\affiliation{Physics Department, Cornell University. Ithaca, NY,
14853}

\date{\today}

\begin{abstract}
We report on new instabilities of the quasi-static equilibrium of
water drops pinned by a hydrophobic inclined substrate. The
contact line of a statically pinned drop exhibits three
transitions of partial depinning: depinning of the advancing and
receding parts of the contact line and depinning of the entire
contact line leading to the drop's translational motion. We find a
region of parameters where the classical
Macdougall-Ockrent-Frenkel approach fails to estimate the critical
volume of the statically pinned inclined drop.
\end{abstract}

\pacs{????}

\maketitle

\section{Introduction}

Dispense a drop on a flat substrate and then tilt it. Depending on
the balance between gravitational and capillary pinning forces,
the drop will slide down or stay at rest. Raindrops sticking or
sliding on a vehicle windshield provide a familiar example of this
drop stability problem, which is of broad practical importance.

In structural genomics, for example, protein crystals are grown by
dispensing protein-containing drops onto horizontal glass or
plastic substrates. Because protein crystals are extremely
fragile, the substrate is then inverted so as to prevent any
nucleating crystals from sedimenting onto and adhering to it
\cite{Hampton,McPherson}. Crystals in these "hanging drops"
instead sediment to the liquid-air interface, where they can be
easily extracted without damage. It was shown previously
\cite{Berejnov} that pinning conditions are important for
maintaining stability of the inclined drops of protein solutions
with concentration suitable for crystallization. Obviously,
crystal nucleation and growth  - and thus the ultimate quality of
the molecular structure determined by X-ray crystallography - are
strongly affected by the drop shape and consequently by the drop
stability. Drop shape variations are also a major obstacle to
automated optical recognition of the protein drop's contents,
important in high-throughput experiments. More generally, the
motion of contact lines is related to motion of elastic manifolds
in the presence of disorder \cite{Fisher}, including motion of
interfaces in porous media and depinning of flux line lattices
\cite{Blatter}, Wigner crystals and charge-density waves
\cite{Thorne}.

\begin{figure}[h]
\includegraphics[width=6.5cm]{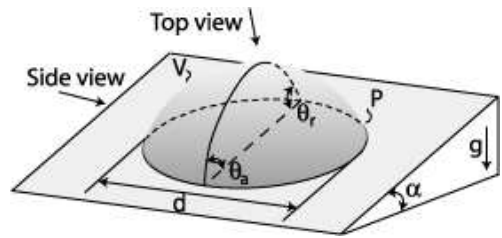}
\caption{\label{fig:fig1} A drop on an inclined surface,
characterized by the drop volume $V$ and diameter $d$, the contact
line perimeter $p$, the advancing and receding contact angles
$\theta_a$ and $\theta_r$, and the substrate inclination angle
$\alpha$.}
\end{figure}

The pinning of an inclined drop that prevents its continuous
motion is related to the contact angle hysteresis, whose magnitude
is usually estimated from the maximum difference between the
contact angles  $\theta_a$ and $\theta_r$ at the advancing
(downhill) and receding (uphill) edges of the contact line, as
shown in Fig.~\ref{fig:fig1}. If this maximum difference
$\Delta(cos \theta)_{r,a}=\cos \theta_r -\cos \theta_a $ is
nonzero, then drops of volume less than a critical $V_c(\alpha)$
may remain at rest at a given inclination angle $\alpha$
\cite{Frenkel, Macdougall}, although this is a necessary but not
sufficient condition for drop stability. A simple formula

\begin{equation}
\Delta(\cos\theta)_{r,a}=A V \sin\alpha \label{eq:eq1}
\end{equation}

was obtained by Macdougall and Ockrent as a phenomenological
explanation of their experiments with the inclined drops
\cite{Macdougall} and independently by Frenkel as a boundary
condition of one exactly formulated problem of the drop stability
on a tilted surface \cite{Frenkel}. Here $A=d^{-1}a^{-2}$ is an
appropriately scaled material constant, $a=(\gamma/\rho g)^{1/2}$
is the capillary length, and $\rho$, $\gamma$, $g$, and $d$ are
the density, surface tension, gravity, and drop width,
respectively.

The MOF formula Eq.~(\ref{eq:eq1}) is believed to describe the
relation between contact angle hysteresis and the equilibrium and
criticality of an inclined drop\cite{Frenkel, Macdougall, Aron,
Bikerman, Kawasaki, Furmidge, Extrand, Roura}, from small
inclinations and volumes at which the drop deforms but remains
static up to the critical inclination or volume at which it begins
to slide continuously. The validity of the MOF functional form has
been verified in a variety of experiments \cite{Macdougall, Aron,
Bikerman, Kawasaki, Furmidge, Extrand, Roura}. However, surprisely
it was noted that all phenomenological improvements have led to
the same general form of Eq.~(\ref{eq:eq1}), with different length
factors instead of $d$  and this formula has not always correctly
predicted the exact volumes of drops pinned at critical
conditions.

Despite extensive study, the problem of the stability of a
one-component drop on an inclined surface has yet to be addressed
in its full richness. Here we examine the equilibrium and
criticality of a water drop pinned on a hydrophobic flat glass
slide. In particular, we examine a wide range of quasi-static
tilting $0^\circ-90^\circ$ for a moderately hydrophobic surface
($\sim90^\circ$ contact angle) having a moderate range of contact
angle differences $\sim30^\circ-90^\circ$. We show that the MOF
criterion is not general and its limitation is based on the
existence of transient modes of displacement, whose mechanics have
not yet been described. We have clearly resolved the partial and
global instabilities of the contact line. Three coexistence curves
corresponding to the partial depinning of the advancing and
receding parts of the contact line and to depinning and continuous
motion of the entire contact line mark those instabilities
unambiguously.  The transient displacements of the advancing and
receding parts of the contact line have different scalings. We
find that the difference between the drop's maximum and minimum
contact angles cannot reliably be used to predict the maximum
pinning strength and the onset of drop sliding.

\section{Contact line stability and depinning for an inclined drop}

A drop's contact line, between its initially static pinned state
and its steady sliding motion, may exhibit a continuous series of
intermediate states. This fact has been previously noted
\cite{Bikerman, Furmidge, Extrand, Extrand2}, but its effect on
the stability of an inclined drop has not been fully appreciated.

After being dispensed on a homogeneous, flat, horizontal
substrate, a drop will have a circular contact line. As the
substrate angle $\alpha$ is slowly stepped upward, a drop of
volume $V$ eventually becomes unstable and slides continuously
down the substrate. At a smaller angle, the drop's contact line
may become locally unstable, and undergo local displacements that
change the contact line's shape but that do not produce continuous
motion. Although these critical angles for the onset of global and
local instability are in general different, for some experimental
conditions they may be weakly distinguishable in measurements. In
this case, the transient pinning of the contact line does not
affect  drop criticality and may be neglected. The drop
equilibrium and criticality may then be described in terms of a
simple energetic balance, as in the MOF formula, between the
potential and capillary energies of the inclined drop. When the
global and local equilibriums  are well separated, the contact
line can be displaced over the substrate while simultaneously
maintaining its  stability against continuous sliding. In this
case, because the contact line configuration corresponding to the
global equilibrium is unknown, the simple energetic balance
describing the drop stability criterion should be reconsidered.

\begin{figure}[h]
\includegraphics[width=7cm]{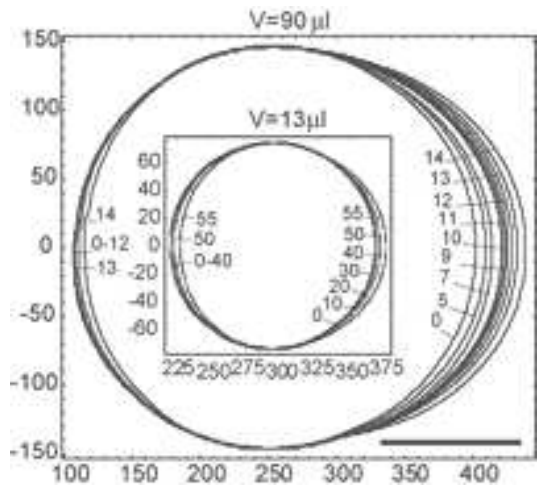}
\caption{Views normal to the substrate plane of the static contact
lines for two drops, achieved as the substrate inclination is
increased. The coordinates are in pixels, with the same image
magnification for both drops, and the scale bar is 2.6 mm long.
The numbers on each contact line indicate the substrate
inclination a in degrees.}\label{fig:fig2}
\end{figure}

Several attempts to analytically describe the shape,
reconfigurations and criticality of the contact line for an
inclined drop have been reported. In 1948, using a variational
technique to analyze the drop shape for small inclinations,
Frenkel \cite{Frenkel} explicitly showed for a 2D inclined drop
that equilibrium conditions and  translational drop instability
lead to the MOF criteria Eq.~(\ref{eq:eq1}). Popova \cite{Popova}
extended the variational technique to a 3D drop at small
inclination. She analytically calculated the equilibrium drop
shape, contact line shape, and the contact angle as a function of
position along the contact line. Carre and Shanahan
\cite{CareeShanahan} used the ideas similar to Popova's variation
of the contact angle along the contact line to calculate the
pinning force, and obtained a criticality equation similar to the
MOF criterion Eq.~(\ref{eq:eq1}). Dussan \cite{Dussan}, stimulated
by earlier experiments \cite{Bikerman, Furmidge}, studied
criticality of 3D inclined drops with an initially
elliptically-like contact line. Using the equations of continuous
fluid dynamics, she found an equation of equilibrium states for
the inclined drop for small hysteresis, which can be used for
obtaining the critical conditions of the inclined drops also.
Popov \cite{Popov} used a variational analysis to examine the
equilibrium and criticality of a weakly perturbed hemi-spherical
drop at large inclinations. His solution describes well only the
stability of small drops inclined near $\alpha=90^\circ$.

It is important to note that in all the above studies
\cite{Frenkel, Popova, CareeShanahan, Dussan, Popov}, the contact
line shape was either assumed arbitrarily or else determined from
the minimum of some free energy.  Furthermore, in all cases the
chosen contact line shape was assumed to be maintained up to and
including the critical point for the onset of  instability and
drop sliding. Consequently, these approaches ignored the
transitional behavior of the contact line prior to global
criticality.

In addition to these analytic attempts, two numerical studies
\cite{Dimitrakop, Iliev}
\begin{figure}[h]
\includegraphics[width=8cm]{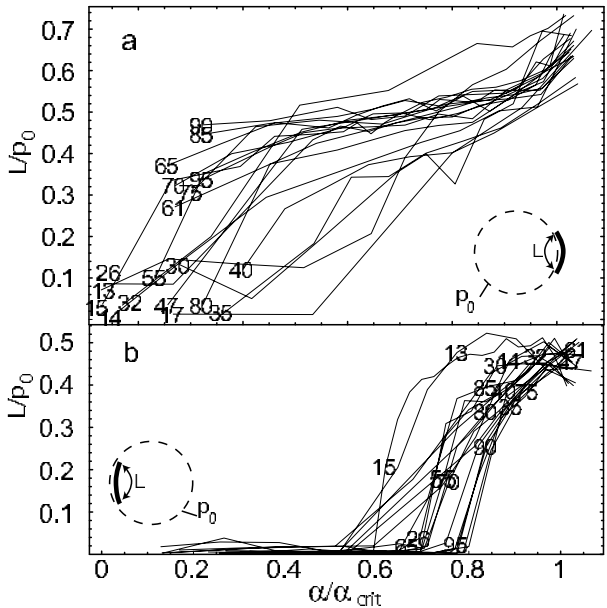}
\caption{Total displaced length of the contact line $L$ at (a) the
advancing and (b) the receding edges as a function of drop
inclination angle $\alpha$. The initial drop perimeter at
$\alpha=0^\circ$ is $p_0$, and at $\alpha_{crit}$ the contact line
can no longer find a static configuration and begins to slide
continuously. The numbers on each curve denote the drop volumes in
$\mu l$.}\label{fig:fig3}
\end{figure}
have explored freely displacing contact lines. Dimitrakopoulos and
Higdon \cite{Dimitrakop} reported transient contact line behavior
similar to that observed here. However, the implementation of
pinning in the numerical algorithm, which leads to the
experimental-like contact line profiles, was arbitrary, and their
$y$-constrained boundary conditions need further clarification and
justification. Iliev \cite{Iliev} used two phenomenological
parameters to describe the pinning. However, the unclear
connection between those parameters and experiment and the absence
of numerically calculated the drop deformations and criticality
curves make comparison with the present results difficult.

\section{Material and experimental methods}

Distilled and deionized water purified by a NANOpure II system
(Barnstead, Boston, MA) was dispensed onto siliconized flat glass
slides with diameter 22 mm (HR3-231, Hampton Research, Laguna
Niguel, CA). On a freshly unpackaged slide, a 40 $\mu l$ water
drop formed a reproducible contact angle of $90-92^\circ$. To
determine drop stability on a given slide, a drop was manually
dispensed onto a horizontal slide using a 100 $\mu l$ micropipette
(Pipetman Co., France). The slide was then slowly rotated in
$2-4^\circ$ steps on a home-built rotation stage. The time
interval between rotations was roughly one minute, long enough to
allow transient shape relaxations to dissipate. A 640x480 pixel
resolution digital camera (Cohu, San Diego, CA) with a telecentric
55 mm objective (Computar, Japan) was mounted on the rotation
stage. Image recording at six frames per second began immediately
after each stage rotation was completed and continued throughout
the entire relaxation period. A custom image recognition program
was written and implemented in LabView to process each image to
extract the contact line. Fig.~\ref{fig:fig2} shows examples of
contact line profiles at different tilt angles determined in this
way.

The apparent contact angles at the advancing and receding contact
lines were measured from the drop side view using an independent
horizontal goniometer. Dispensed drop volumes were accurate to
$0.1-0.5\%$, and tilt and contact angle measurements were accurate
to $1-2^\circ$. Measured velocities $U$ of average drop contact
line motion of the transient displacements were $<0.1$ mm/s. Using
water's dynamic viscosity $\eta\sim0.01$ g/(cm s) and surface
tension $\gamma\sim70$ dyn/cm yields an upper bound $< 10^{-6}$
for the capillary number $Ca=U\eta/\gamma$. Thus all dynamic
effects during contact line rearrangement can be neglected.

On a horizontal, homogeneous flat surface, a drop's minimum
free-energy configuration has a circular contact line. In
practice, the actual contact line shape depends on the initial
contact conditions formed while the drop is dispensed. We found
that the subsequent contact line displacements depend on the
contact line's initial shape and on the initial contact angles
along it. Consequently, we carefully prepared and selected drops
with initially circular contact lines.

\section{Results}

Fig.~\ref{fig:fig2} shows typical results for the contact line
position recorded at different tilt angles, for two drops with
volumes of 13 $\mu l$ and 90 $\mu l$, respectively. As the
substrate is tilted, the contact line remains pinned in its
original circular configuration. Beyond a first critical tilt
angle $\alpha_a$, the advancing portion of the contact line
becomes locally unstable and displaces in an attempt to find a new
equilibrium, eventually reaching a new static configuration.
Beyond a second, larger critical angle $\alpha_r$, the receding
part of the contact line becomes locally unstable and displaces,
but the drop again finds a new static configuration. Beyond a
third critical angle of inclination $\alpha_{crit}$, the drop
becomes unstable and slides continuously. The difference in the
behavior of the advancing and receding contact lines implies that
pinning along the contact line is not homogeneous. This conclusion
is consistent with previous calculations and measurements
\cite{Finn, Brown, CareeShanahan, ElSherbini} of the contact
angles along a contact line's circumference.

Fig.~\ref{fig:fig3} examines how the displaced length $L$ of the
static contact line evolves with respect to the critical
parameters $\alpha$ and $V$. The total displaced length $L$ of the
contact line at the advancing and receding edges following a tilt
increment to angle $\alpha$ can be determined by subtracting
images acquired at $\alpha=0^\circ$ from those at that angle
$\alpha$.
\begin{figure}[h]
\includegraphics[width=8.5cm]{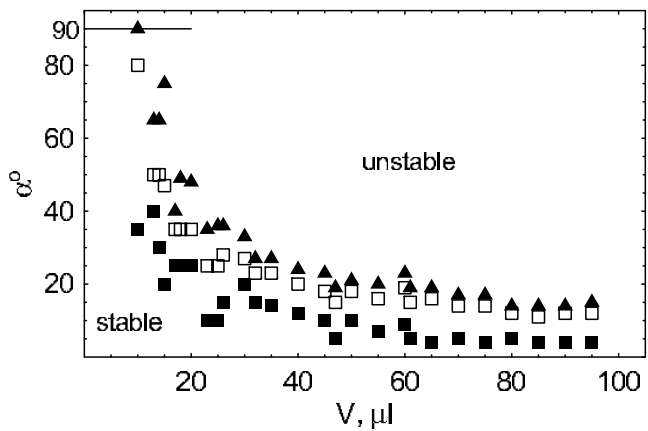}
\caption{\label{fig:fig4}Contact line transitions as a function of
substrate inclination angle $\alpha$ and drop volume $V$. Symbols
denote $\blacksquare$: onset of quasi-static displacements of the
advancing contact line, $\square$: onset of quasi-static
displacements of the receding part, and $\blacktriangle$: onset of
continuous motion of the whole contact line.}
\end{figure}
For small $\alpha$ such that the contact line remains
circular, the resultant images still show some displaced pixels
that are randomly distributed over the contact line, arising from
noise and other measurement errors. In Fig.~\ref{fig:fig3} (a) and
(b), this regime corresponds to the horizontal "zero" parts of the
curves. At larger $\alpha$ (beyond $\alpha_a$ or $\alpha_r$), the
advancing or receding part of the contact line begins to displace
(Fig.~\ref{fig:fig3} insets), and the difference image shows a
chain of connected pixels. This chain grows on further inclination
to form a displacing front of length $L$.

Fig.~\ref{fig:fig3} shows the ratio of the total displaced length
$L$ to the unperturbed drop perimeter $p_0$ (at $\alpha=0^\circ$)
versus $\alpha/\alpha_{crit}$, where $\alpha_{crit}$ corresponds
to onset of drop instability and continuous translational motion.
Fig.~\ref{fig:fig3} visualizes unambiguously the fact noted
previously \cite{Bikerman, Furmidge, Extrand, Extrand2} that the
advancing contact line may displace at a lower inclination angle
than the receding contact line. Although there is considerable
scatter in the data, there is still remarkable consistency in
behavior over the factor-of-10 volume range examined. The
advancing contact line begins moving at a small
$\alpha/\alpha_{crit}<0.2$ and the displaced length $L/p_0$ grows
monotonically, reaching a consistent value of $\sim0.6\pm0.1$ just
before $\alpha_{crit}$. In contrast, the receding contact line
remains pinned until $\alpha/\alpha_{crit}>0.6-0.8$, and then
steeply increases to $L/p_0\sim0.5$ at $\alpha_{crit}$. Note that,
because the drop becomes distended, the total contact line length
near $\alpha_{crit}$ exceeds its initial length $p_0$.

\begin{figure}[h]
\includegraphics[width=8.5cm]{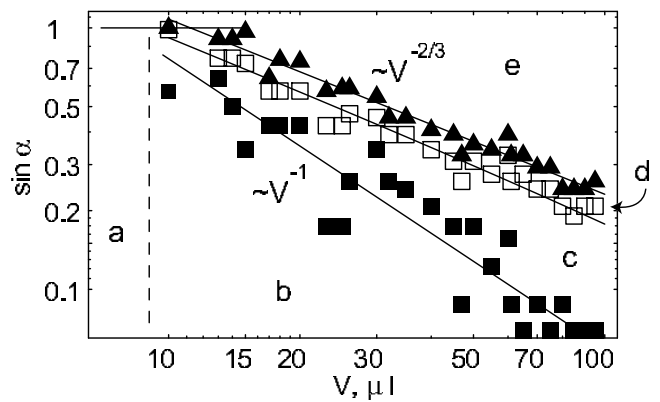}
\caption{Log-log representation of Fig.~\ref{fig:fig4}. The
letters denote zones where (a), the contact line is absolutely
stable against any inclination; (b) the contact line is stable up
to a maximum inclination $<90^\circ$; (c) the contact line locally
displaces to a new static configuration; (d) the contact is
globally displaces to a new static configuration; (e) the contact
line is unstable and moves continuously.}\label{fig:fig5}
\end{figure}

Fig.~\ref{fig:fig4} presents a subset of the data in
Fig.~\ref{fig:fig2}, plotted in the space of critical parameters
$V$ and $\alpha$. The solid $\blacksquare$ and open $\square$
squares indicate the onset of local instability at the advancing
and receding parts of the contact lines, respectively, and the
solid triangles $\blacktriangle$ indicate the onset of continuous
drop motion. A log-log scaling of this data is presented in
Fig.~\ref{fig:fig5}, and clarifies the observed differences in the
transient displacements of the contact lines.
Between absolute stability (zone (a)) - where the initial circular
contact line is maintained at all inclinations - and continuous
motion (zone (e)), the contact line passes through three
transitions: instability of the advancing line at $\alpha_a(V)$
(Fig.~\ref{fig:fig4}, curve $\blacksquare$), instability of the
receding line at $\alpha_r(V)$ (Fig.~\ref{fig:fig4}, curve
$\square$), and finally, instability at $\alpha_{crit}(V)$ leading
to continuous translational motion of the entire contact line
(curve $\blacktriangle$). Fig.~\ref{fig:fig5} also clearly shows
that there are five zones of behavior in the space $(V, \alpha)$
in which the contact line loses its stability. In addition to
zones (a) and (e), in zone (b) the contact line is stable only up
to a maximum inclination $<90^\circ$. In zone (c), between the
$\blacksquare$ and $\square$ transition curves, the contact line
shows partial instability at its advancing edge. In zone (d) both
the advancing and receding portions participate in quasi-static
displacements, but the drop still remains at rest.

It is interesting to note the functional dependencies of the
transitional curves. The receding contact line transition and the
transition to continuous sliding have the same volume scaling
$V^{-0.75}\sim V^{-2/3}$. However, the advancing transition scales
as $V^{-1.06}\sim V^{-1}$. This suggests that the advancing
instability is controlled by disturbances that do not scale like
the drop size. The length factor $d$ in the MOF formula for the
advancing part of the contact line should thus be replaced by some
new scaling $\delta$, which is independent of the drop size and is
likely related to the length scale of local perturbations.

\begin{figure}[h]
\includegraphics[width=8.5cm]{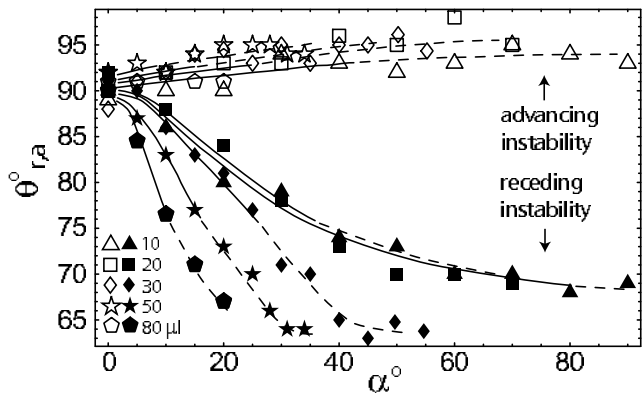}
\caption{Apparent values of the advancing contact angle $\theta_a$
(open symbols) and receding contact angle $\theta_r$ (closed
symbols) vs. inclination $\alpha$, for five different volumes $V$.
The lines are guides to the eye, and are comprised of a solid
segment where the contact line remains pinned in its initial
$\alpha=0^{\circ}$ configuration (regions (a) and (b) in
Fig.~\ref{fig:fig5}) and a dashed segment where the contact line
undergoes local displacements (regions (c) and (d) in
Fig.~\ref{fig:fig5}). The contact line as a whole remains in
static equilibrium until the end of the dashed line, and slides
continuously beyond it (the $\blacktriangle$-line in
Fig.~\ref{fig:fig5}).}\label{fig:fig6}
\end{figure}

Fig.~\ref{fig:fig6}  shows the measured apparent advancing and
receding contact angles versus inclination angle, for drop volumes
ranging over a factor of 8. The measurements were performed for
each particular drop volume at inclinations below $\alpha_{crit}$,
for which the contact line reached a static configuration after
each inclination increment. Solid guide lines denote the stable
regions ((a) and (b) in Fig.~\ref{fig:fig5}), and dashed lines
indicate zones of partial instability at the advancing and
receding edges ((c) and (d) in Fig.~\ref{fig:fig5}). The drop
contact line traverses the stable zones (a) and (b), passes the
advancing and receding displacement transitions and the zones of
partial instability (c) and (d), and eventually reaches the
absolutely unstable zone (e). These data deviate significantly
from those presented in Ref. \cite{Wolfram}, where a hydrophobic
substrate was also used. In particular, while the advancing angle
in Fig.~\ref{fig:fig6} is nearly independent of drop volume, the
receding angle depends strongly on drop volume.

Fig.~\ref{fig:fig6} raises another interesting question, touched
on in \cite{Krasov}: which values of the advancing and receding
contact angles $\theta_a$ and $\theta_r$ for a given drop volume
do we have to choose for an adequate description of contact line
pinning and stability? According to Fig.~\ref{fig:fig5} and
Fig.~\ref{fig:fig6}, the advancing and receding angles often lie
in different zones of stability in Fig.~\ref{fig:fig5}.  In
particular, the last absolutely stable of $\theta_r$ (indicated in
Fig.~\ref{fig:fig6} by the points at which the solid curves
connect to the dashed ones) do not have corresponding absolutely
stable values of $\theta_a$; the advancing line has undergone
quasi-static displacements that modify (reduce) $\theta_a$ from
what would be obtained if the advancing line had remained in its
initial $\alpha=0^\circ$ position. In this case when the angles
lie in different zones of stability, they cannot be simply
inserted into the MOF formula to get an accurate measure of
pinning strength and drop stability. Traditionally, the contact
angle hysteresis at criticality is obtained from the difference
$\theta_a - \theta_r$ measured at the last inclination before the
drop begins to slide. These correspond to the last points on the
dashed curves in Fig.~\ref{fig:fig6}, and to the $\blacktriangle$
- transition curve in Fig.~\ref{fig:fig4} and Fig.~\ref{fig:fig5}.

\begin{figure}[h]
\includegraphics[width=8.5cm]{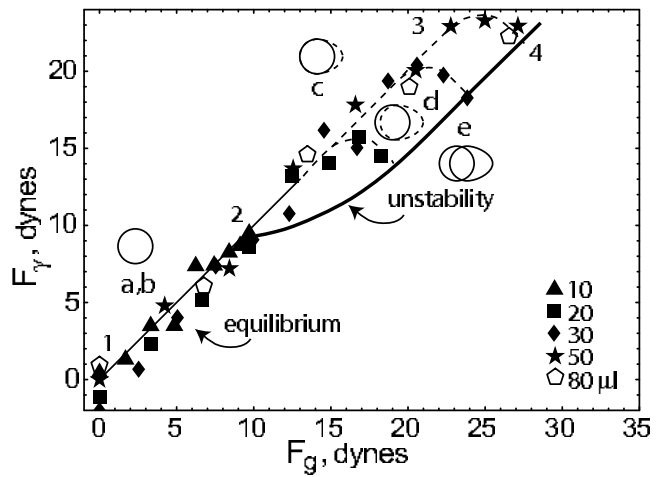}
\caption{Equilibrium and criticality of an inclined drop. Symbols
indicate the evolution of drops from static pinning to continuous
sliding, for five different volumes. The solid line 1-2 is a fit
to the MOF formula (Eq.~(\ref{eq:eq1})), the dashed lines show how
the data deviates from this fit and the solid curve 2-4 indicates
instability of the contact line (the $\blacktriangle$-line in
Fig.~\ref{fig:fig5}). The letters and inserts represent the same
regions plotted in Fig.~\ref{fig:fig5} and different kinds of
displacement of the contact line respectively.  For definitions of
$F_{\gamma}$ and $F_g$ see the text.}\label{fig:fig7}
\end{figure}

Fig.~\ref{fig:fig7} shows a direct test of the MOF equation as a
predictor of the  equilibrium and stability of inclined drops.
Using the perimeters $p$ from images similar to those presented in
Fig.~\ref{fig:fig2} and the traditional choice for the contact
angles $\theta_a$ and $\theta_r$ to quantify the  contact angle
difference, we may rescale the data of Fig.~\ref{fig:fig4} and
Fig.~\ref{fig:fig6} using $F_{\gamma}=\gamma(\cos \theta_r  - \cos
\theta_a)p/2\pi$ and $F_g=V \rho g \sin \alpha $ such that the MOF
fit of Eq.~(\ref{eq:eq1}) appears as a straight line. At small
values of the scaled variables, the data are in fact linear, and a
fit gives $\gamma=72$ dynes/cm, consistent with the accepted value
for water at T=22°C of 72.5 dynes/cm. The part 1-2 of the solid
line corresponds to the regions (a) and (b) in
Fig.~\ref{fig:fig5}, where the contact line of pinned drops is in
equilibrium and does not change its initial configuration (see an
insert corresponding to (a, b)-region in Fig.~\ref{fig:fig7}).
Thus, the MOF approach is able to describe well deformation of a
meniscus upon the drop inclination while the contact line remains
completely pinned and the gravity and capillarity forces can
balance each another. This behavior in this region of parameter
space has been observed in many previous studies \cite{Macdougall,
Bikerman, Kawasaki, Furmidge, Extrand, Roura}.

After the point 2 at larger $F_g$ and $F_{\gamma}$, corresponding
to larger inclinations or drop volumes or contact angle
differences $\theta_a-\theta_r$, the equilibrium curves deviate
substantially from the linear fit: line 2-3 (dashed curves in
Fig.~\ref{fig:fig7}). The line 2-3 corresponds to the region (c)
in Fig.~\ref{fig:fig5} where the advancing part of the contact
line is not stable and exhibits quasi-static displacements (see an
insert corresponding to the region (c) in Fig.~\ref{fig:fig7}).
The points at largest $F_g$ - on the "hooks" of the dashed curves
in Fig.~\ref{fig:fig7} - represent the last stable contact line
configurations, corresponding to the $\blacktriangle$ -line in
Fig.~\ref{fig:fig5}. The zone between the curves 2-3 and 2-4
corresponds to the (d) region of Fig.~\ref{fig:fig5}. According to
Fig.~\ref{fig:fig5},  in this zone both the advancing and receding
contact lines are unstable but only undergo quasi-static
displacements (not sliding).  These last points form the
criticality curve 2-4, which obviously, cannot be described by the
MOF formula.

\section{Conclusions}

The present experiments show that the standard picture of
instability of inclined drops is over-simplified, and that the
behavior is more complex than assumed in previous analytical
treatments. In particular, the assumption that the contact line
remains unchanged with inclination until it begins to slide does
not adequately describe the actual depinning.

As the substrate tilt is increased, three distinct depinning
transitions - of the advancing portion of the contact line, the
receding portion of the contact line, and of the contact line as a
whole - are observed. These transitions have different effects on
overall drop stability.  Transient displacements of the advancing
contact line in zone (c) decrease drop stability locally
($\blacksquare$-curve in Fig.~\ref{fig:fig4}).  But on further
inclination, in zone (d), transient displacements of the receding
contact line (occurring with additional advancing displacements)
involve whole contact line in reconfiguration that stabilizes  the
overall pinning ($\blacktriangle$-curve in Fig.~\ref{fig:fig4} and
the "hooks" in Fig.~\ref{fig:fig7}).  We understand this
stabilizing effect as arising a new mechanism of dissipation
directed against the drop sliding, which origin is based on
pinning-depinning of the local parts of the drop contact line.

Consistent with the MOF formula, the contact angle difference
$\Delta(\cos\theta)_{r,a}$  varies linearly with $\sin \alpha $ up
to inclinations at which the receding contact line displaces.  But
the MOF formula fails to describe drop stability at larger angles
and the critical transition to drop sliding.

\begin{acknowledgments}
This work was supported by the NIH (R01 GM65981-01). We thank C.
W. Extrand for helpful and stimulating e-mail discussion and P.
Dimitrakopoulos and M. H. Murray for critical reading of our
manuscript.
\end{acknowledgments}

\end{document}